\documentclass[]{aastex63}
\usepackage{listings}
\usepackage{amsmath}
\usepackage{bm}
\usepackage{url}

\makeatletter
\def\@head{head}
\makeatother

\shorttitle{Algorithmic Pulsar Timing}
\shortauthors{C. Phillips}

\graphicspath{{./}{figures/}}

\begin{document}

\title{Algorithmic Pulsar Timing}

\author[0000-0002-2099-0254]{Camryn Phillips}
\email{clp3ef@virginia.edu}

\affiliation{University of Virginia, 530 McCormick Rd., Charlottesville, VA 22904, USA}

\author[0000-0001-5799-9714]{Scott Ransom}
\email{sransom@nrao.edu}

\affiliation{National Radio Astronomy Observatory, 520 Edgemont Rd., Charlottesville, VA 22903, USA}

\begin{abstract}
Pulsar timing is a process of iteratively fitting pulse arrival times to constrain the spindown, astrometric, and possibly binary parameters of a pulsar, by enforcing integer numbers of pulsar rotations between the arrival times.  Phase connection is the process of unambiguously determining those rotation numbers between the times of arrival (TOAs) while determining a pulsar timing solution. Pulsar timing currently requires a manual process of step-by-step phase connection performed by individuals. In an effort to quantify and streamline this process, we created the Algorithmic Pulsar Timer, APT\footnote{APT is available at \url{https://github.com/clp3ef/APT}}, an algorithm which can accurately phase connect and time isolated pulsars. Using the statistical F-test and knowledge of parameter uncertainties and covariances, the algorithm decides what new data to include in a fit, when to add additional timing parameters, and which model to attempt in subsequent iterations. Using these tools, the algorithm can phase-connect timing data that previously required substantial manual effort. We tested the algorithm on 100 simulated systems, with a 99\% success rate. APT combines statistical tests and techniques with a logical decision-making process, very similar to the manual one used by pulsar astronomers for decades, and some computational brute-force, to automate the often tricky process of isolated pulsar phase connection, setting the foundation for automated fitting of binary pulsar systems. 
\end{abstract}

\section{Introduction} \label{sec:intro}
 Pulsars are neutron stars that spin rapidly and emit beams of electromagnetic radiation from their magnetic poles. The magnetic poles' misalignment with the rotation axis of the neutron star causes the beams to sweep across the sky \citep{essentialradio}. When a pulsar’s beam sweeps across the Earth, we observe a regular series of flashes, much like a lighthouse. Pulsar timing is the process of unambiguously accounting for every rotation of a pulsar over an extended period of time. Because pulsars are internally stable, they spin at extremely regular rates that afford pulsar timing a high level of precision. Many millisecond pulsar spin periods can be measured to twelve or more significant figures. At this precision, effects that are negligible for other astronomical observations, such as orbiting planets or relativistic effects within the Solar System, become highly relevant  \citep{backer_hellings_1986}. Pulsar timing measures these effects through the precision counting of pulses via a series of iterative fits until all known physical effects have been accounted. 
 
\begin{figure}[ht]
\includegraphics[scale=0.38]{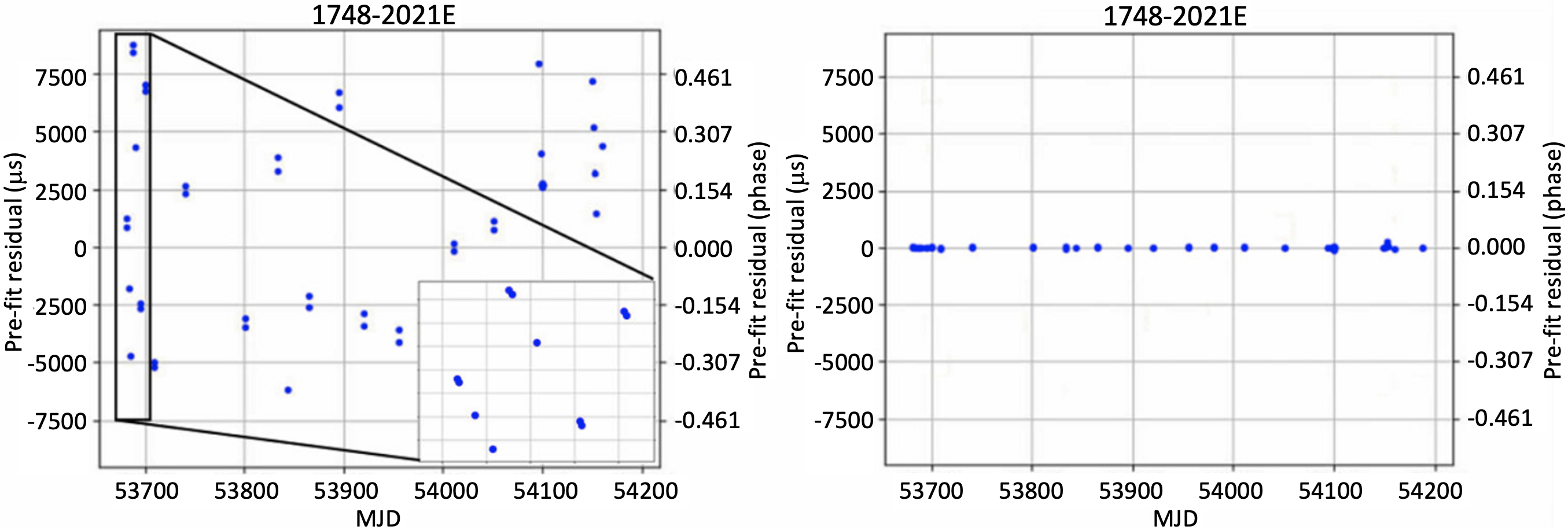}
\caption{Timing residuals (i.e. data minus model) plots for PSR J1748$-$2021E, before (left) and after (right) determination of a timing solution. Pulsar timing is the process of turning the left plot into the right by doing a series of iterative fits with increasing spans of data and progressively more complete model parameters. The left plot is the system state after initial discovery plus monitoring observations. This allows for a rough position estimate based on survey data and the low accuracy spin found from the search mode data. We simulate this initial discovery state for large scale testing as described in \S\ref{subsec:data}. The errors on the pulse times of arrival (TOAs) are similar to the size of the points at this scale.} \label{fig:1}
\end{figure}

Figure~\ref{fig:1} shows two timing residual (i.e. observation minus model) plots of isolated pulsar PSR J1748$-$2021E, before and after being timed. The y-axes of these plots are in pulse phase and time, where pulse phase is defined from $-$0.5 to $+$0.5 of a rotation.  In a residual plot, the best fit model is subtracted from the measured arrival times or TOAs. If the model perfectly accounts for all effects present, this produces flat residuals, as seen on the right. Unmodeled effects will produce systematics in the residuals, such as a sinusoid for a position error. Most physical effects observed in pulsar timing produce unique residuals; the four most important are spin frequency $f$ (or $F0$ as denoted by pulsar timing software) with a linear signal, Right Ascension and Declination with sinusoids of period one year, but different amplitudes and phase, and spin frequency derivative $\dot{f}$ (or $F1$) with a quadratic \citep[see][Fig 6.7]{essentialradio}. A timing model consisting of only a roughly known spin frequency and nothing else (which may be better described as a \emph{lack} of a timing model) produces a plot like the one on the left of Figure~\ref{fig:1}: pulse arrival times are consistent within small groups, but lack any clear connection between observations. Normality of TOA errors is given by the work of \cite{2015}, who show in their Appendix B that TOA errors are very close to Gaussian so long as the signal to noise ratio is greater than or about equal to 10, a reasonable assumption for most pulsars.

Because of the overlapping complexity of multiple fit parameters with unique signals, pulse number ambiguities between TOAs, and inconsistently and/or inadequately sampled data, often with large gaps where hundreds of thousands or even many millions of rotations of the pulsar were unobserved, a computer cannot perform a simple curve fit to the data using brute force; the algorithmic complexity scales exponentially with the number of TOAs. Statistical methods have been used on a few specific pulsars to determine phase connection between TOAs \citep[e.g.][]{2008ApJ...673L.163V, 2009ApJ...692L..62K}, but never as a fully automated pulsar timing pipeline, though \textsc{DRACULA} and \textsc{COBRA}, described below, have both begun the process of automating the pulsar timing pipeline. 

The recent package \textsc{DRACULA}\footnote{\url{https://github.com/pfreire163/Dracula}} by \citet{2018MNRAS.476.4794F} is a brute-force phase connector, which uses the fact that the reduced $\chi^2$ for a fit diverges quadratically when incorrect numbers of phase wraps occur between TOAs. While \textsc{DRACULA} currently requires TOAs to be specifically formatted with JUMPs for the program, it is a very useful method. We implement a rudimentary version of \cite{2018MNRAS.476.4794F}'s algorithm, described in \S\ref{subsec:algorithm}.

The \textsc{COBRA}\footnote{\url{https://github.com/LindleyLentati/Cobra}} code described in \cite{2018MNRAS.473.5026L} is a Bayesian framework which builds an effective timing model consisting of spin, dispersion measure, and binary parameters via coherently combining pulsed signals from multiple search-mode data sets. \textsc{COBRA} does not fit for sky position, assumes only Gaussian pulse profiles, and is very computationally expensive, making it a prime candidate for use in conjunction with APT. It might be possible for future versions of COBRA to determine what are effectively arrival times of averaged pulse shapes which could be used by APT to phase-connect over much larger amounts of data, or perhaps portions of data with large gaps in time. In addition, binary parameter fits found from \textsc{COBRA} can be frozen into input models to allow for the fit of binary systems by APT, as described in section~\ref{sec:improve}. \textsc{COBRA} is currently unmaintained but still serves as a proof of concept of the automation of pulsar timing.

Despite these advancements, the vast majority of pulsar timing has been done by hand, using a series of rules and recipes passed down from mentors to students. We attempted to encode those rules into an algorithm which could emulate the decision-making process of a human timing a pulsar. Two main tools aid this decision-making process: predictive models based on the full covariance matrix of the previous fit, and the F-test to judge whether adding a fit parameter significantly improves the fit or not.  Both are implemented through \textsc{PINT}, a modular, Python-based, pulsar timing software package \citep{luo2020pint}\footnote{\url{https://github.com/nanograv/PINT}}. While the methodology we present is generalizable, APT requires and relies heavily upon \textsc{PINT} scripts, methods, and classes.

\section{Predictive Models with Gaussian Error Estimates} \label{sec:models}

The first step for an automated timing algorithm is deciding, given a current best fit model to a subset of the data, which more-complex model to attempt next, or alternatively, should we use the same model with more data?  This decision can be simplified further: given a subset of fitted data and a new (unfitted) data point, can the residual of the new datum be accounted for by varying the current fit parameters, or is a new model parameter needed? One way to decide is to slightly perturb the current model based on the correlated errors and parameter values from the current fit and observe whether the perturbations account for the residuals of the new data points. Such a perturbation can be produced as follows.

Given a single parameter and its associated error, one can produce a Gaussian distribution around the best fit value with the error determining the width of the distribution. The single-variable normal distribution of a timing parameter and its associated error can be generalized to a multi-variable normal distribution if we use the full covariance matrix of the fitted parameters. The 1 by $d$ mean vector $\boldsymbol{\mu}$ and the $d$ by $d$ covariance matrix $\boldsymbol{\Sigma}$, where $d$ is the number of fit parameters, take the places of the mean and variance in a single-variable normal distribution,
\begin{equation} \label{eq:1}
\boldsymbol{y} = f(\boldsymbol{x},\boldsymbol{\mu},\boldsymbol{\Sigma}) =
\frac{1}{\sqrt{\boldsymbol{\Sigma}(2\pi)^{d}}}\exp(-\frac{1}{2}(\boldsymbol{x}-
\boldsymbol{\mu})\boldsymbol{\Sigma}^{-1}(\boldsymbol{x}-\boldsymbol{\mu})^{'}).
\end{equation}
Equation~\ref{eq:1} is the probability distribution function of the $d$-dimension multivariate normal distribution, where $\boldsymbol{x}$ is a 1 by $d$ vector representing the independent variables and $\boldsymbol{y}$ is a 1 by $d$ vector representing the resulting probabilities for each possible value of $\boldsymbol{x}$. The mean vector contains the best fit values of all the parameters, while the diagonal elements of the full covariance matrix are the square of the errors of each fit parameter and the off-diagonal elements are the covariances between fit parameters. We use both the diagonal and off-diagonal elements to construct the Gaussian distributions. The post-fit covariance matrix is provided by a class within \textsc{PINT}, and the multivariate Gaussian distribution is implemented using the \texttt{NumPy.random} module. See \S\ref{subsec:code} for further details. 

\begin{figure}[ht]
\includegraphics[scale=0.43]{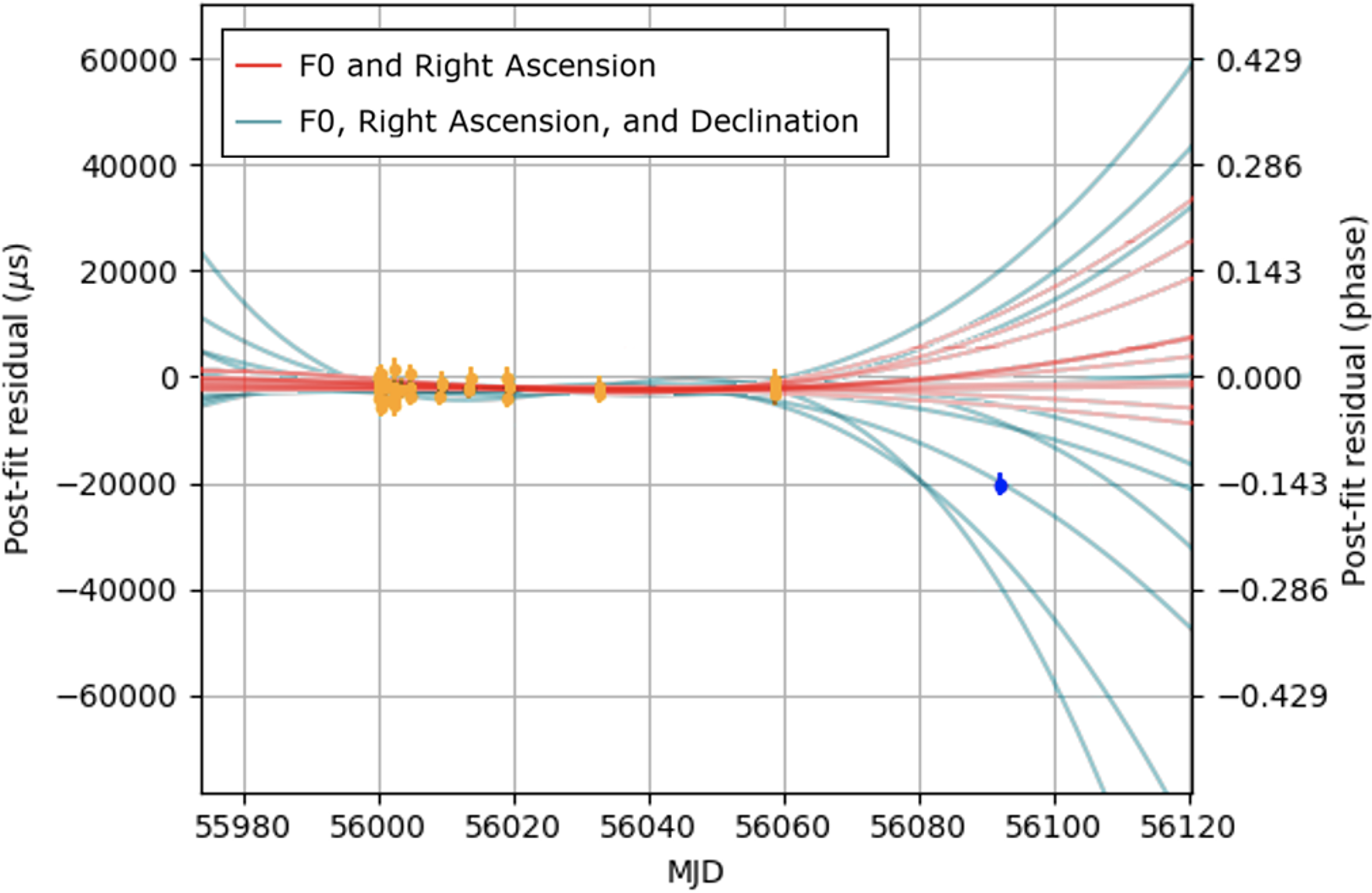}
\caption{An overlay of two residual plots during phase connection. The yellow points have been fit with parameters $F0$ (i.e.~spin frequency, $f$) and Right Ascension, while the blue point is next to be fit. The red predictive models show that the blue points are \emph{not} within the span of the current model parameters and cannot safely be included using only the current fit parameters. This tells the user that an additional parameter is needed, most likely Declination. The cyan predictive models show the result of adding Declination. The blue point is now within the span of the models, telling the user that adding Declination can account for the residuals of the new point, so Declination should be included in the next model. \label{fig:3}}
\end{figure}

By randomly selecting parameter values from the multivariate Gaussians created by Equation~\ref{eq:1}, a perturbed model is produced with parameter values that, while not optimal, are still consistent with the data. If many such perturbed models are plotted, each with parameters randomly chosen from their respective Gaussian distributions and including covariances, they agree strongly at the points included in the current fit but vary widely over the extrapolated times beyond the fit points. Figure~\ref{fig:3} shows an example of such predictive models.
 
The lines answer the question of which model to try next. If new data are within the spread of the predictive models then it is possible to account for the residuals of those points by varying the existing parameters. If the points are significantly outside the spread, a new parameter is likely needed. Also importantly, whichever line passes closest to the next data point is a better model including those data than the current model, and therefore is used as the starting model for a subsequent fit.

\section{F-test} \label{sec:F-test}
The F-test is a simple statistical measurement that is widely used in data modelling to compare two nested models with different numbers of degrees of freedom. Models with more degrees of freedom inherently fit data better, but it may be that the improvement due to an additional degree of freedom is due to chance rather than actually needing additional fit parameters. The F-test assesses this by comparing $\chi^2$ values and numbers of degrees of freedom for both models to calculate the probability that the fit improvement might be due to random chance. By comparing the current best fit model to a copy of itself with an additional fit parameter, the F-test tells us whether a new parameter is needed, or whether the current parameters can account for the current residuals. 

\section{Algorithm} \label{subsec:algorithm}
Using predictive models and the F-test, we created an algorithm called APT to determine a phase-connected timing solution for isolated pulsars. We omit binary pulsars systems because they require a separate algorithm and are a minority in the pulsar population, only constituting around one in ten known pulsars\footnote{See ATNF pulsar catalog at http://www.atnf.csiro.au/research/pulsar/psrcat/}. The manual process of fitting an isolated pulsar is relatively simple: starting with at least two TOAs, the most prominent parameter (typically the spin frequency, $F0$) is fit to those TOAs, a decision of the goodness-of-fit is made, and either additional data or new model parameters are added. The process repeats until all TOAs have been included and a timing solution with $F0$, the spin frequency derivative $F1$, and sky position (typically Right Ascension and Declination) have been fit. APT replicates this process using statistical tests in place of human judgement, and Figure~\ref{flowchart} in the appendix shows the structure of the algorithm's logic as a flowchart. 

The program begins by evaluating the local number density of TOAs, $n_{x}$, at the position of every TOA, as shown in Equation~\ref{eq:4}, where $t_i$ is the arrival time in MJD of the i-th TOA and $t_x$ is the arrival time in MJD of the TOA being scored, where i evaluates over every TOA other than x. This creates a ranked list where TOAs with many nearby observations are ranked highest, because in those regions, the number of unknown rotations between TOAs can be better constrained. APT iterates through the highest scoring starting points until either a valid solution for the pulsar is found, it runs out of possible starting points, or it reaches the optional cap on iterations, which defaults to five to prevent run time being wasted on badly scoring TOAs. 

\begin{equation} \label{eq:4}
n_{x} = \sum_{i=1}^{N_{\rm TOAs}} \frac{1}{\mid{t_{i}-t_{x}\mid{}}} \;\;\;\;\; (i \neq x)
\end{equation}

After determining the starting points for the attempt, the main loop of the program begins with three objects: a current best fit model, which might simply be a starting frequency and astrometric position from a pulsar search observation, the subset of TOAs which is being fit, and a boolean array that selects subsets of TOAs from the full set of TOAs to be fit. In this first iteration, the TOA subset is the highest scoring TOA from the list described above, plus its closest neighboring TOA. 
	
The starting model is fit to the current TOA subset, providing a baseline fit. Using this baseline and the initial TOA subset, twelve\footnote{This number is adjustable by the user but defaults to twelve, which is the value that will be used in the text for simplicity.} predictive models are produced from the covariance matrix of the best fit model (see Equation~\ref{eq:1}). Using the current fit TOAs plus the closest group of TOAs, the $\chi^2$ of the residuals is calculated for all twelve predictive models plus the baseline model. A "group" is a set of TOAs collected in the same observation. A group is typically phase-connected within the group, where "phase-connected" means the exact number of pulsar rotations between any two TOAs is known. The algorithm chooses the model with the smallest $\chi^2$ as the new best fit model, and the closest group of TOAs is appended to the current TOA subset. If the optional polynomial extrapolation or bad point checking features, described below, are invoked, they operate here.

It is valid to ask why the algorithm doesn't simply \emph{fit} the next data points instead of using the twelve predictive models and choosing the closest. The reason is that the predictive models help bridge the gap between the current best fit model and the model that would be produced by simply fitting. Since the predictive models, by definition, must be within the bounds of the parameter errors, using them adds an extra buffer to prevent the algorithm from performing a fit which would move the current parameters far from their best fit values and  sharply increase their errors.

APT performs an F-test for each possible additional parameter to determine if a new parameter should be added to the model. The possible parameters are Right Ascension, Declination, and spindown rate ($\dot{F}$, or $F1$), and each has a minimum time span that the current TOA subset must exceed before that parameter can be added. This prevents parameters that act on long timescales from being added to the model before they could realistically have an effect. The parameter with the smallest probability of the F-test statistic that is also less than an adjustable probability limit (default $P < 0.0005$) is added to the best fit model. If none of the F-test results are less than this limit, or no parameters can be added because the time span is not long enough yet, the best fit model retains the same number of parameters. 

At this point in the program, the TOA subset has been expanded by at least one group of phase-connected TOAs to create a new TOA subset, and there is a new best fit model, possibly with an additional parameter. However, if there was a large enough gap between the fit TOAs and the nearest group, or if the fit was particularly poor, as indicated by a relatively large $\chi^2$ value, it is possible that the newly appended group was phase-wrapped, meaning that it was assigned to an incorrect rotation of the pulsar. The true residual phase of a wrapped TOA is above +0.5 or below -0.5 phase, but is wrapped back into the range of -0.5 to +0.5 because of the fact that pulsar timing assumes integer numbers of rotations between TOAs, and so fits operate modulo those integers on the fractional phase residuals. Another way to see it is that there is an ambiguity in the number of pulses which have arrived between one observation and the next. You don't know if a positive residual means, for instance, that the current pulse arrived a bit late for the model in question, or if the $next$ pulse in the model arrived too early. The only way to determine if a phase wrap is present is to attempt a fit with the phase wrap included.

We implemented a rudimentary form of the phase-wrap checking process as described by \cite{2018MNRAS.476.4794F}. In order to bridge gaps in TOAs and prevent the assigning of an incorrect number of rotations between TOAs, the algorithm performs all the same calculations as above with the closest group of TOAs, but also assuming $n$ additional rotations between those TOAs and the earlier TOA subset, for a total of 2$n$ + 1 (+ 1 being the $n$ = 0 case) fits being performed. Each trial produces a best fit model with or without new parameters according to individual F-tests, and the algorithm compares the final $\chi^2$ values for each of the 2$n$+1 trials. Whichever trial produces the smallest $\chi^2$ is accepted as the true best fit model, and the rest are abandoned. 

As stated, this is a rudimentary version of the algorithm proposed by \citet{2018MNRAS.476.4794F}, since it does not attempt to find the minimum $\chi^2$ using the quadratic relation between the reduced $\chi^2$ and the phase wrap \citep[see][Fig 5]{2018MNRAS.476.4794F}. Our algorithm explores phase wrapped solutions on gaps with residual differences greater than a minimum threshold (default 0.15 in phase), and by default it only attempts phase wraps $n$ in the range -1 to +1, although these default guidelines can be overridden by the user in order to utilize the algorithm's computing power to test large numbers of phase wraps much more quickly and efficiently than a person could. The simplicity of the defaults is largely due to the presence of the predictive models which solve TOA gaps via random model extrapolation rather than brute force. A full implementation of the phase-wrapping algorithm described by \citet{2018MNRAS.476.4794F} is a possible improvement to APT, but the current method is a sufficient first-order approximation given the overwhelming majority of pulsar timing does not require phase wraps larger than a few rotations.

Once the model with the smallest $\chi^2$ has been chosen, the algorithm continues on with the new best fit model and extended subset of TOAs, possibly with a correction for a phase wrap. The best fit is plotted, saved as an image, and the new TOAs, model, and selection array are saved so a user may later reload data from any point in the fit. The main loop of the program begins again with the new best fit model and extended TOAs as the starting model and data, and this process continues until all TOAs have been included. The program ends when the reduced $\chi^2$ at the end of a run is below the cutoff value (defaults to 10). The default is set higher than 1 since we are more concerned with collecting all correct answers than accidentally including incorrect answers, and poor fits have reduced $\chi^2$ on the order of thousands, well above this threshold. If the reduced $\chi^2$ at the end of a run is above the cutoff, the process starts again with the next pair of starting points. This continues until APT runs out of starting points or finds a successful fit.

\subsection{Optional Features} \label{subsec:features}

The algorithm has two optional features that can be activated from the command line: polynomial extrapolation and bad point checking. Both improve specific parts of the algorithm, but also have weaknesses which lead to their inclusion as being optional.

\begin{figure}[ht]
\includegraphics[scale=0.5]{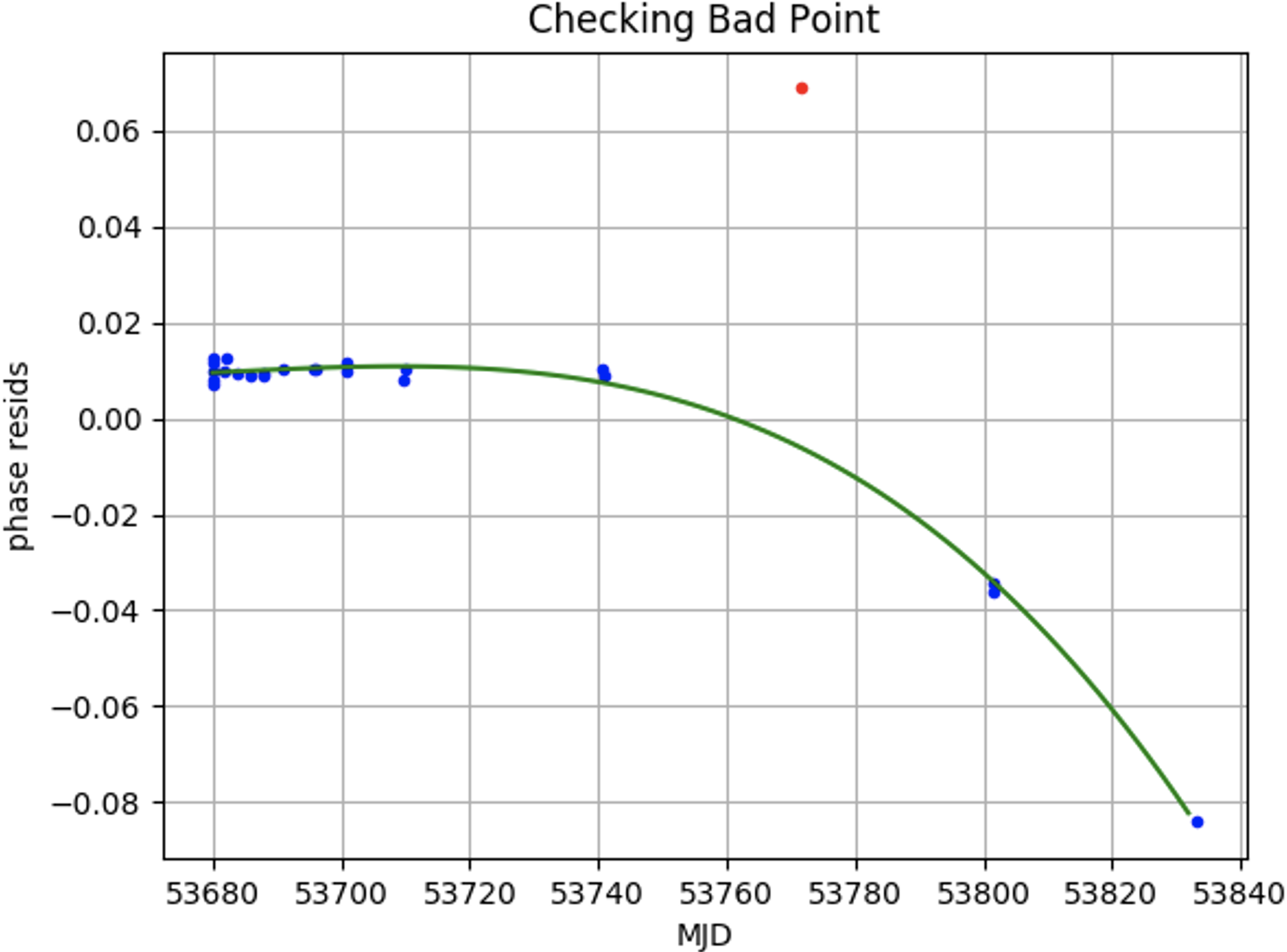}
\caption{Bad group checking. The algorithm is questioning the red group, and is checking its validity by fitting a polynomial to the three groups after it. The success of the polyfit strongly implies the red group is bad, and it will be ignored for the remainder of fitting.  \label{badtoa}}
\end{figure}

In manual pulsar timing, multiple new TOA epochs are often added at once when their predicted residuals seem to follow a smooth curve. This was replicated in the algorithm with the polynomial extrapolation option, which, when invoked, fits a third-order unweighted polynomial using \textsc{NumPy.polyfit} to all data within 1.3 times the current fit span. If the square sum of the residuals of the polynomial fit is less than a set threshold (default 0.02 phase), the algorithm attempts to fit a polynomial on a span 1.8 times and then 2.4 times larger than the fit span\footnote{These span multipliers are variable and the values listed are the defaults.}, and appends all TOAs from the largest successful fit to the current fit TOAs, rather than only appending a single phase-connected group. This option significantly speeds up the timing process, often reducing run times by 50 to 60 percent. However, it also poses a danger since it may incorrectly include bad TOAs if they appear to be in-line with predictions by chance. If you know your data may have bad TOAs and APT is consistently failing, it may be necessary to try a run with polynomial extrapolation turned off.

If a pulsar is particularly faint, or if there are problems with interference, instrumentation, or software, it is possible for TOAs to be in error and have incorrect phases. In manual fitting, such points are deleted or ignored when identified. For APT, we hesitated to make this a mandatory feature since, for especially difficult timing problems, the algorithm could latch onto a wrong initial ``solution'', making later good data appear bad and causing it to ignore good data. The balance we found is checking for bad points only when the phase wrap check is activated and allowing the user to toggle this feature. The phase wrap check activates when the next TOA has a residual, compared to the currently fit TOAs, greater than 0.15 phase, which may suggest that TOA or group of TOAs is bad if it is not phase wrapped. APT checks the validity of the group by temporarily ignoring it and fitting a polynomial to the next three groups of TOAs. This will fail in most cases since residuals greater than 0.15 suggest the next several TOAs are phase-wrapped and unconnected, as one would expect. However, if the polynomial fit is good, this strongly implies the to-be-fit group is significantly out of line with the rest of the data and should be ignored in subsequent fits. Figure~\ref{badtoa} shows an example of this decision being made.  This check is designed to ignore groups only if it is extremely clear they are bad, and leaves more subtle data vetting to the user. 

\section{Simulated Data} \label{subsec:data}
\begin{figure}[ht]
\includegraphics[scale=0.54]{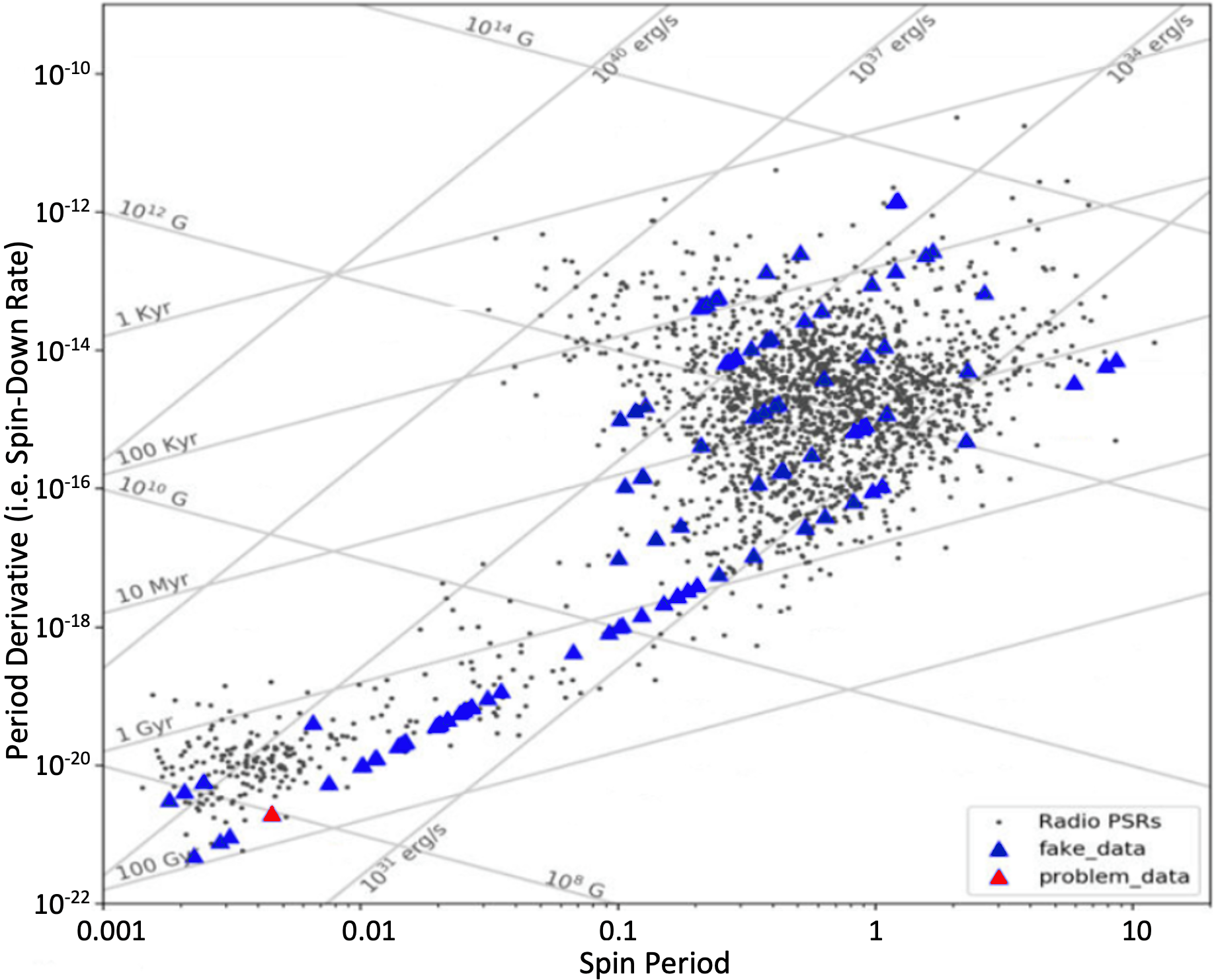}
\caption{The 100 simulated pulsars, marked by blue-edged triangles, displayed over the traditional $P-\dot{P}$ diagram. The simulated systems sample several pulsar subclasses, including millisecond pulsars (lower left), ``normal'' radio pulsars (middle), young and energetic pulsars (upper middle), and magnetars (upper right), showing the versatility of the algorithm. The pink-centered triangle is the single unsolved system. \label{ppdot}}
\end{figure}

In order to robustly test the algorithm, we simulated pulsar systems and then randomly skewed the starting parameter estimates to prevent our knowledge of the solution influencing our judgement of the algorithm's performance. We used the algorithm to attempt to solve these systems blindly, and in fact did not know whether the number of TOAs and their separation would allow an unambiguous solution. We produced 100 of these systems, 30 with initial frequencies of millisecond or recycled pulsars (10\,Hz $< f <$ 800\,Hz) and 70 with initial frequencies of slow pulsars (3\,Hz $< f <$ 10\,Hz). Figure~\ref{ppdot} shows the distribution of the simulated pulsars as compared to known pulsars in $P-\dot P$ space. We drew the parameters from realistic ranges based on the known relation between spin and spindown rate, if we assume a constant magnetic field strength. We chose the simulated pulsar positions randomly over the entire sky. The simulated data was in the form of TOAs rather than search mode data since APT is not designed to discover or time pulsars directly in search data, but to create basic timing models from TOAs from folded observations for known but not yet phase connected pulsars.  

We simulated the TOAs using \textsc{zima}, a simple TOA simulation code within \textsc{PINT}. Our simulation code, which uses \textsc{zima} internally, intentionally emulates typical timing campaigns for newly discovered pulsars. The general guideline is as follows: two observations occur within the first 24 hours, then observations occur three of the next five days, then two of the next ten days, then one in the next week, and then monthly until the end of the data, which is at minimum 200 days and at maximum 700 days. To ensure substantial variation while still following this guideline, these were interpreted as number densities of observations for each time span described. The exact number of observations and separation between consecutive observations were allowed to vary uniformly within each time span such that, on average, the number of observations within a time span divided by the sum of the separations between those observations reproduced the number densities described by the guideline. The uniform ranges were chosen so that the standard deviations of the resulting number densities were $\sim 1$, with greater variation allowed in the later time densities. 

Each observation is assigned a random integer number of TOAs between 1 and 8, with a special exception for the first two observations, which can have from 3 to 8 TOAs. This additional sampling at the beginning is to prevent phase wraps within the first day of observations and to emulate the typically longer duration observations that occur at the beginning of real timing campaigns. All TOAs within a simulated pulsar dataset have the same error, randomly chosen from the uniform range 0.0003 to 0.03 in phase, which is then scaled to the appropriate time error using the spin rate. Again, these observations are sets of TOAs, not raw search mode data.

Once the pulsar's timing parameters are chosen and TOAs are simulated, we estimate the much lower-precision starting timing parameters for the pulsar, including a rough spin rate, Right Ascension, and Declination. This is consistent with real survey data where the spin rate and position are only approximations and all other parameters are completely unknown. Note that dispersion measure is typically determined with enough precision during the search observations that it can be fixed at the search value during phase connection, unless the observing frequencies for the timing observations are very different than those of the search observations\footnote{For a good example of this practice, see the determination of Dispersion Measure in \citet{article}}. For the pulsar position, we randomly shift the ``observe'' position from the true position by a normal distribution whose standard deviation is a FWHM chosen from a uniform range of 2 to 40 arcminutes, based on typical uncertainties from radio pulsar survey detections. 

We ``blur'' the spin rate by randomly choosing a phase difference in the range 0.05 to 0.15 phase and dividing by the length of the observations, representing a conservative error how well the spin frequency can be measured during a search observation. The resulting offset is added to the known spin rate value, and the parameter uncertainty adjusted accordingly. Finally, we set the spindown rate to zero since it cannot be determined before long-duration timing observations are complete. 

The system parameters now exactly resemble the level of precision seen in survey data for unsolved pulsars. The new set of parameters with realistic errors is used as the starting model for the algorithm, which reconstructs the correct parameters using pulsar timing as described in the previous section. At the beginning of the timing process, spin is the only parameter being fit, consistent with real pulsar data where spin is the first and most important constraining parameter. Because the parameter uncertainties are unused in the fitting process, and will be recomputed during the first fit, they are arbitrarily set to 10$^{-2}$ for the positional parameters, 10$^{-6}$ for spin rate, and zero for spindown rate.  The script which produces these simulated isolated pulsar systems, called \texttt{ simdata.py}, is available as part of APT.

\section{Results} \label{subsec:results}
Before we produced simulated data, we tested the algorithm on and successfully fit two real isolated pulsar systems, J1748$-$2021E and J1748$-$2446G, with final timing solutions identical to those produced by a human fit of the same data. The algorithm was told to start with the first two TOAs, but was otherwise left to run uninterrupted. These systems are typical millisecond pulsars and prove APT's ability to solve real pulsars using real observations.

We used APT to fit 100 simulated systems, and it successfully solved 99 of them. Table~\ref{table:results} summarizes the process of solving the 100 systems, including the options used for each round of testing. All simulations and testing were performed on a 2016 ASUS Notebook running Windows 10 with an Intel Core i5-6200U CPU.

\begin{deluxetable}{cc}[ht]
\tablenum{1}
\tablecaption{Number of Systems Solved Given Listed Input}
\label{table:results}
\tablewidth{0pt}
\tablehead{\colhead{Input} & \colhead{Number of Systems Solved}}
\startdata
No Additional Input & 75 \\
Set Starting Points to First Observation &	85 \\
Shorten \texttt{RAJ} and \texttt{DECJ} Minimum Fit Spans & 89 \\
Adjust \texttt{F1} Minimum Fit Span & 99 \\
\enddata
\end{deluxetable}

APT solved 75 of the 100 systems with no user intervention. Given the TOA data and rough estimates of the spin rate and position, the algorithm chose a starting TOA, iterated through the TOAs, added parameters as necessary, and, within an average run time of 4 minutes on an Intel Core i5-6200U CPU, phase connected the pulsar. In addition, phase wraps were necessary for the solutions of 18 of the 100 systems, displaying the efficacy and importance of checking for phase wraps. When using APT to solve a pulsar, there are several indications that a fit is going well: the median $\chi^2$ values printed in each iteration stay below 100, phase wraps are rarely if ever checked, and polynomial extrapolation is activated several times. Large median $\chi^2$ vales, the constant testing of phase wraps, and a significantly slower run time as polynomial extrapolation fails, are all indications that a run is going poorly and is unlikely to produce a phase-connected solution. 

In our testing, 25 of the 100 systems showed these signs of failure, and were only able to produce a phase connected solution with user intervention. Ten of these failures were due to poor choice of starting TOA based on the method described in \S{\ref{subsec:algorithm}}. Equation~\ref{eq:4} presents the metric by which starting points are chosen, and shows its main flaw: because number density is evaluated on a TOA by TOA basis, a single observation containing many closely spaced TOAs may be favored over several closely spaced but short observations. This sometimes causes APT to begin the phase connection process at a suboptimal location. APT faces this problem in two ways: by continuing to try to solve the pulsar with different starting TOAs in every run, and by allowing the user to specify a starting TOA. It is completely possible there are more efficient formulas than the one described by Equation~\ref{eq:4}. However, so long as the formula used prioritizes density of observations over density of TOAs within observations, it is sufficient for our purposes.

Fourteen of the  fifteen remaining failures were due to the order in which parameters were added. For systems with particularly large positional uncertainties (i.e.~greater than 8 arcminutes), the minimum span before Right Ascension and Declination could be added may need to be shortened so that the position can be fit for more quickly than the defaults allow for. Setting the minimum \texttt{ RAJ} span to 1.5 days and the minimum \texttt{ DECJ} span to 2.0 days allowed for an additional 5 systems to be solved. It is possible, especially for young pulsars (including magnetar-like systems), that the spindown rate becomes significant sooner than the Right Ascension, and needs to be added before either positional parameter. APT accounts for this by adjusting the minimum time span for $F1$ to be added according to the estimated spin rate, so that pulsars with $F0 < 10$\,Hz can start fitting for $F1$ after 9.7 days, while pulsars with $10 < F0 < 100$\,Hz can start fitting for $F1$ after 30.6 days, and for $F0 > 100$\,Hz, APT must wait 97 days. However, it is still possible that $F1$ needs to be added sooner or later than this estimated span. In three of the remaining ten failing cases, adjusting the minimum time span for $F1$ to four days allowed APT to solve the system successfully. It is also possible that, for systems with particularly stable spindown rates, that the default minimum span is too short, allowing $F1$ to be fit for before Right Ascension or Declination. For six of the ten systems, adjusting $F1$'s minimum span to 150 days allowed for the positional arguments to be determined first and the systems to be solved. 

The single remaining unsolved system was a millisecond pulsar with particularly skewed initial positions and large error bars. APT was successfully able to fit for spin and Right Ascension in the first week of observations. However, at 15 days there is an observation consisting of a single data point, and the next observation is not for another 14 days. This scarcity of data implies possible phase wraps in the gaps between observations, but phase wraps from $-$12 to 12 were tested and did not solve the system at this decision point. The minimum parameter spans for Declination and spindown were adjusted to force spindown to fit before Declination; both parameter addition orders produced identical failure-bound fits within a few iterations. Increasing the F-test cutoff to allow for more flexibility in adding parameters similarly had small effects on parameter addition timings but did not produce successful results. 

In order to identify the source of difficulty, each parameter value was individually adjusted towards the solution value until, with all other parameters unchanged, APT could solve the system. Adjustment of spindown led to the fastest results, with only small adjustments towards the solution needed to allow APT to connect across the larger gaps in data. Adjustments of Right Ascension and Declination had minimal effect, requiring the values be set to nearly their solution values before a solution was found. Additionally, the adjustment of spin had little effect at any level, as the fit conducted by APT within the first few days accurately approximate the true value closely enough that setting spin to the solution value had no significant effect. These tests show that poor position estimates amplify the effects of poor spindown estimates, such that a poor spindown fit can turn a difficult system into an unsolvable system if the system is already highly skewed in position (greater than 10 arcminutes). Additionally, the failing system was particularly data sparse, with several observations containing a single TOA and many large gaps between consecutive observations. The addition of two observations between 15 and 29 days provided a connection across the largest gap and allowed spindown to be fit with a very reasonable F-test much sooner than without those points. This system provides an excellent example of the issues that may face scientist and algorithm alike - poor data sampling, inaccurate estimates, and large error bars.

While APT is able to successfully fit 99 of the 100 simulated systems and two real isolated pulsars, J1748$-$2021E and J1748$-$2446G, there are other possible sources of error that a user may encounter. We outline here some known issues that may arise and how to address them. If data has several large gaps which APT is struggling to fit across, try setting the maximum number of wraps it can test to a larger number. If APT is choosing a bad starting point, which is especially likely if there is one day with many observations at multiple frequencies, a user provided starting point will likely work better than the algorithm's first choice. If APT appears to be adding parameters out of the expected order (the title of each saved plot shows which parameters are being fit), try adjusting the minimum parameter spans to control the order or time at which parameters are added. If the user knows there may be data with incorrect phases, such as from a faint or new pulsar, turn off polynomial extrapolation so it does not accidentally include bad data points. Conversely, if user data is known to have no poorly calculated TOAs, but APT is removing data points, turn bad point checking off. APT also does not fit for dispersion measure, so TOAs fed to APT should either be at the same frequency or DM should have been externally fit and frozen into the parfile in order to keep errors in pulse phase much less than one. Nearly every internal parameter and limit of APT can be set by the user on the command line and are described in detail in the code appendix \S\ref{subsec:code}, and we encourage users to explore the code and adjust their input parameters as needed for their particular data.  

\section{Future Improvements} \label{sec:improve}
Although the algorithm is functional, it has significant room for improvement. Currently, the algorithm does not work with JUMPs, fitted 'floating' phase offsets between groups of TOAs which are locally phase-connected. Handling systems with pre-existing JUMPs by removing or ignoring the JUMPs would be a simple first step, since the algorithm is designed to work without them. Restructuring the code to not just tolerate, but actively use, JUMPs would make the algorithm more flexible and effective, but would significantly increase the complexity and likely, the run time of the code.

Implementing JUMPs properly is the first step in allowing APT to solve binary pulsars, as well. Binary pulsars require a completely different approach involving JUMPs around every group of TOAs and allowing for the fitting of orbital parameters. The main loop of a binary fitting algorithm would work to globally improve timing and orbital parameters while iteratively combining JUMP-ed groups of TOAs. Allowing for the identification and resolution of phase wraps within JUMPs would also be a necessary tool for a binary algorithm. As a caveat, APT is able to work with binaries in its current form so long as the binary parameters are pre-fit such that their effects are negligible as compared to the four main parameters, spin, Right Ascension, Declination, and spindown. JUMPs would also need to be turned off or removed for APT to run properly. 

Several improvements could be made to the APT script, in general, rather than the algorithm specifically. Currently, the algorithm takes $\sim$2 seconds per iteration, with several hundred iterations per fit. APT's run time is directly proportional to the number of observations. This makes the program time-consuming, as much of the extra time comes from the predictive models which require the calculation of absolute phase for all TOAs. This calculation is necessary for correct predictive models, but could possibly be optimized to shorten the run time. The program could also be polished in several ways: allowing for other niche parameters, such as proper motion and higher derivatives of spin rate, to be included in the fit once all four main parameters have been solved, and adding more command-line functionality.

\section{Conclusions} \label{sec:conclusions}

We have created a new algorithm, APT, which can phase connect realistic isolated pulsar timing data sets. APT successfully solved two real isolated pulsar systems, J1748$-$2021E and J1748$-$2446G, and 99 of 100 simulated systems. It performs these fits with an average run time of $\sim$4 minutes, and contains several features, including the testing of phase wraps, the removal of bad data, and extrapolating intermediate solutions forward in time. APT is most notable as a proof of concept, similar to \textsc{DRACULA} \citep{2018MNRAS.476.4794F} and \textsc{COBRA} \citep{2018MNRAS.473.5026L}, since pulsar timing is almost universally a manual process. It is possible that this algorithm could be a starting point for further research into algorithmic pulsar timing, especially in solving binary pulsar systems. APT shows that the decisions made by a human in the pulsar timing process can be encoded into a relatively simple algorithm. In the future, such automatic pulsar timing software could save astronomers many hours of work in solving pulsars. 

APT has particular relevance given the completion of the CHIME telescope in 2017. APT's design goal is to produce an initial pulsar timing model for isolated pulsars consisting of the four major parameters, Right Ascension, Declination, spin, and spindown, for a system which is regularly observed for at least a year and has a pre-defined dispersion measure. Dispersion measure is found from frequency-domain  rather than time-domain observations, so it can either be ignored in single-frequency data or fit from an initial mutli-frequency observation.  While further parameters such as parallax and higher order derivatives of spin could be included, the goal is to create a broadly accurate and reasonable model, not to fully determine every detail of the system. We are safe to make this simplification since the excluded parameters are only significant on long timescales ($\sim$10 years), and would be near impossible to parametrize without first solving for position, spin, and spindown \citep{backer_hellings_1986}. CHIME observes all pulsars in the Northern hemisphere with a minimum cadence of 10 days \citep{2018IAUS..337..179N}. This creates an uncommonly large surplus of pulsar data and requires new tools to efficiently process and analyze said data. APT can act as one of these tools by finding basic models for isolated pulsars, freeing up investigators' time to focus on binaries and exotic pulsar systems. Additionally, APT excels at updating existing models with new data, which complements CHIME's frequent and regimented pulsar observation schedule. While APT was not designed specifically for CHIME, the trend of pulsar astronomy towards large data and automation (see section~\ref{sec:intro}) inspired this prototype. It is our hope that the algorithm described above acts not only as a tool for those using data from telescopes like CHIME, but also as a proof of concept and a stepping stone for further inroads into automation of pulsar timing. 

\section{Acknowledgements} \label{sec:acknowledgements}
This project began within the framework of the National Radio Astronomy Observatory Research Experience for Undergraduates, which was funded by the National Science Foundation through AST-1358169.  The National Radio Astronomy Observatory is a facility of the National Science Foundation operated under cooperative agreement by Associated Universities, Inc.  SMR is a CIFAR Fellow and is supported by the NSF Physics Frontiers Center award 1430284.

\software{PINT \citep{luo2020pint}, NumPy \citep{harris2020array}, AstroPy \citep{astropy:2018}}

\clearpage
\bibliographystyle{aasjournal}
\bibliography{APT.bib}

\section{appendix}\label{sec:appendix}
\subsection{Code References}\label{subsec:code}
\texttt{.tim file} - time data files, contain all TOA data of the pulsar, constant throughout algorithm

\texttt{.par file} - parameter file, contains all information about the current model of the system, always changed and updated by the algorithm

\texttt{.csv file} - comma separated value file, used to store a filter of the TOAs currently being fit so that a user may reload from any point in the process

\texttt{fitter.py} - class within PINT, contains all information and functions used for fitting the model to the data. Used to fit the timing model to the current TOA subset and to output the covariance matrix to compute the random models.  

\texttt{multivariate\_normal()} - function in \texttt{ NumPy.random} module which produces multivariate Gaussian distributions. These are used for each parameter to randomize the predictive models while still keeping the predictions statistically consistent with the known parameter values and errors

\texttt{NumPy.polyfit} - module to fit a n-degree polynomial to data by minimising the squared error. Used by the algorithm to A) test for extreme outliers and B) speed up the process by including larger subsets of data when a third order polynomial can clearly model the folded data, implying the data is already very close to the expected model. All instances of \texttt{NumPy.polyfit} in APT use a third order polynomial in order to allow for greater flexibility than a linear or quadratic fit, but not so much flexibility that sparse or unrelated data would be overfit. 

\texttt{zima} - tool in PINT which allows for the creation of regularly spaced TOAs with specified residuals

\texttt{RAJ} - right ascension at J2000, measured in units of hours, minutes, seconds

\texttt{DECJ} - declination at J2000, measured in units of degrees, arcminutes, arcseconds

\texttt{F0} - spin rate in seconds per second

\texttt{F1} - rate of change of F0 in seconds per seconds squared

\texttt{APT} - Algorithmic Pulsar Timer, code developed to allow for the automatic phase connection of isolated pulsars given $\sim$1 year of unconnected TOA data
\\

\texttt{APT.py} - main program of APT, contains the algorithm and is called from command line

\parindent20mm \texttt{Inputs:}

\texttt{parfile}- Mandatory input. Parameter file to read model from, contains the initial guesses for the four main parameters plus any additional parameters. See also \texttt{.par file}

\texttt{timfile} - Mandatory input. Input time data file, contains a list of all the TOAs and their frequency of observation, observation error, and any labels. See also \texttt{.tim file}

\texttt{starting$\_$points} - Optional input. Defaults to TOAs chosen by algorithm using Equation~\ref{eq:1}. Which TOAs the user wants the algorithm to begin the fitting process with. The starting points may be provided by the user as a comma separated pair of numbers, where two floats indicate the start and end of a MJD range and two integers indicate the TOA groups which should be used. Groups are indexed chronologically, so the first observation is always group 0, the second observation group 1, et cetera. Providing the same integer twice indicates that only one group should be used (i.e. 1,1 indicates group 1, the second observation, should be used as the starting subset of TOAs)

\texttt{maskfile} - Optional input, defaults to None. A \texttt{.csv file} containing a one-dimensional array of true/false values with a length equal to the number of TOAs in the system. Can be read in to notate which TOAs are included in a fit and which are not. The algorithm saves a maskfile, timfile, and parfile at the end of every iteration to allow for reloading from any point in the fitting process. See also \texttt{.csv file} 

\texttt{n$\_$pred} - Optional input, defaults to 10. The number of predictive models to be calculated for APT to choose from on each iteration

\texttt{ledge$\_$multiplier/redge$\_$multiplier} - Optional inputs, default to 1.0/3.0, respectively. Scale factor for how far to plot predictive models to the left/right of fit points

\texttt{RAJ$\_$lim} - Optional input, defaults to 1.5 (days). Minimum time span before APT is allowed to fit for Right Ascension (RAJ).

\texttt{DECJ$\_$lim} - Optional input, defaults to 2.0 (days). Minimum time span before APT is allowed to fit for Declination (DECJ). 

\texttt{F1$\_$lim} - Optional input. Minimum time span before APT is allowed to fit for Spindown (F1). Defaults to the time for F1 to change residuals by 0.35 phase, which is calculated based on the input prediction for F0 based on the general ranges of pulsar spindown rates as compared to spin rates, see Figure~\ref{fig:3} 

setting any of the above limits to a different time span can control the order in which APT adds parameters to the model

\texttt{Ftest$\_$lim} - Optional input, defaults to 0.0005. The upper limit for what is considered a "successful" Ftest value in determining whether a parameter should be added or not. F-test ranges from 0 to 1 with 1 being complete numerical insignificance and 0 being absolute necessity. Defaults to 0.0005, but can be set higher or lower as needed to give more or less leeway in adding parameters.

\texttt{check$\_$bad$\_$points} - Optional input, defaults to True. A true/false value telling APT whether you would like it to check for outliers and, if they do not match the trends of the surrounding data, ignore the data for the remainder of the fit. See also \texttt{NumPy.polyfit}

\texttt{plot$\_$bad$\_$points} - Optional input, defaults to False. True/false value telling APT whether it should save plots when it checks if outlier points do not match the rest of the data. 

\texttt{check$\_$bp$\_$min$\_$diff} - Optional input, defaults to 0.15 phase. APT determines outliers based on the phase difference between the most recently fit TOA and the next TOA to be fit. This input sets the minimum phase difference these two TOAs must present for APT to check if it is an outlier.

\texttt{check$\_$bp$\_$max$\_$resid} - Optional input, defaults to 0.001. The bad point check works by ignoring the next group to be included and fitting a polynomial to a few groups following the excluded group. If the polynomial fit's residuals are below this input value, implying that the questionable group is an outlier, the group in question is ignored. If the polynomial residuals are greater than this input value, APT continues as normal with the questionable group included. 

\texttt{check$\_$bp$\_$n$\_$groups} - Optional input, defaults to 3. The number of groups past the questionable group to fit the bad point check polynomial to. Too far and a phase wrap may be reached and imply a discontinuity where on does not exists. Too short and any trends showing a continuity that the group in question obscures may not be accurately identified.

\texttt{try$\_$poly$\_$extrap} - Optional input, defaults to True. True/false value telling whether to try to speed up the fitting process by including groups which follow a clear trend and are able to be fit by a third degree polynomial. See also \texttt{NumPy.polyfit}

\texttt{plot$\_$poly$\_$extrap} - Optional input, defaults to False. True/false value telling whether to plot the polynomial fits during the extrapolation attempts. This will interrupt the program and require manual closing. 

\texttt{pe$\_$min$\_$span} - Optional input, defaults to 30 (days). Minimum span in days before APT is allowed to attempt polynomial extrapolation. Too soon will lead to meaningless fits as fitting only one or a few points will always have extremely small residuals, whether or not they truly show a trend.

\texttt{pe$\_$max$\_$resid} - Optional input, defaults to 0.02. Maximum acceptable goodness-of-fit for \texttt{NumPy.polyfit} to allow the polynomial extrapolation to succeed and move on to the next span or to include all the currently tested points. 

\texttt{span1$\_$c/span2$\_$c/span3$\_$c} - Optional input, defaults are 1.3, 1.8, 2.4, respectively. Coefficient for first/second/third polynomial extrapolation span (i.e. try \texttt{NumPy.polyfit} on current span multiplied by span1/2/3$\_$c.) Each span is larger than the last, and this continues until the third level is reached or the extrapolation no longer works, at which point all groups already included are added to the to-be-fit TOA subset and included in the current fit.

\texttt{max$\_$wrap} - Optional input, defaults to 1. How many phase wraps APT will test in each direction, such that a max$\_$wrap of 3 means APT will test all phase wraps from -3 to 3. 

\texttt{plot$\_$final} - Optional input, defaults to True. True/false value telling whether to plot the final residuals at the end of each attempt. This will show whether APT was successful in finding a solution or not. Contributes to the run time calculation at the end of the program. 
\\

\parindent4mm
\texttt{simdata.py} - function in APT which randomizes parameter and TOA data to create realistic simulated data for isolated pulsars. Produces three files, a \texttt{.sol file} which is a parameter file containing the "true" parameter values of the pulsar, a .\texttt{.par file} containing the skewed parameters which simulate the inaccuracy of parameter estimates when the pulsar is newly discovered, and a \texttt{.tim file} containing the simulated TOAs to go with the above models. 
\parindent20mm

\texttt{Inputs:}

\texttt{iter} - Optional input, defaults to 1. The number of independent pulsar systems the code will produce. 

\texttt{name} - Optional input, defaults to fake$\_\{$number$\}$, where $\{$number$\}$ is the next largest system number in the \texttt{fake$\_$data} folder, assuming the folder only contains files of the above naming convention. If no such folder exists, \texttt{simdata.py} will create it and the first system will be named \texttt{fake$\_$1}. If input is given, output files will be of the format $\{$name$\}$.par, etcetera. Providing a custom name only works with \texttt{iter} $=$ 1

\texttt{F0$\_$value/RAJ$\_$value/DECJ$\_$value/F1$\_$value} - Optional inputs, default to None. Sets F0 (Hz)/ RAJ (degrees)/ DECJ(degrees)/F1 (1/$s^2$) to the given value, respective to each input parameter. For example, RAJ$\_$value $=$ 3.4 will set the right ascension of the simulated pulsar to 3.4 degrees, and F1$\_$value $=$ 1.2e$-$16 will set the spindown of the simulated pulsar to 1.2e$-$16 1/$s^2$. This overwrites the randomization based on realistic ranges and distributions of each parameter that is defaulted to, and this input can only be used when \texttt{iter} $=$ 1

\texttt{F0$\_$error/RAJ$\_$error/DECJ$\_$error/F1$\_$error} - Optional input, all default to 0.0000000001. Set value for the parameter error of F0 (Hz)/ RAJ (degrees)/ DECJ(degrees)/F1 (1/$s^2$), respectively. Default error value does not matter as errors will be recalculated with the first fit, meaning the input errors have no effect on APT. This input can only be used when \texttt{iter} $=$ 1

\texttt{f0blur/rblur/dblur/f1blur} - Optional inputs, default to values randomly sampled from appropriate distributions as described in \S\ref{subsec:data}. The amount by which to skew the known value of F0 (Hz)/ RAJ (degrees)/DECJ(degrees)/F1 (1/s$^2$), respectively, such that the resulting \texttt{parfile} is a reasonable simulation of a starting model for a newly discovered pulsar. These inputs can only be used when \texttt{iter} $=$ 1

\texttt{f0blur$\_$range} - Optional input, defaults to '0.05, 0.15' phase, where the comma separated pair is a string that will be interpreted in the code as a tuple. The range of uniform random phases to skew F0 by (phase), set so that the spin cannot be skewed so harshly that phase wraps exist within the first group of TOAs. 

\texttt{PEPOCH} - Optional input. Period epoch for pulsar in MJD, should match TZRMJD. Defaults to 56000 so as to be in recent years for which asterometric data is available. 

\texttt{TZRFRQ} - Optional input. Frequency (Hz) at which the pulsar was observed, defaults to 1400 Hz

\texttt{TZRMJD} -  Optional input. Day the first TOA will be set to (MJD), defaults to 56000

\texttt{TZRSITE} - Optional input. Observation site code, defaults to GBT (Green Bank Telescope) 

\texttt{density$\_$range} - Optional input, defaults to '0.004, 0.02' days/TOA, where the comma separated pair is a string that will be interpreted in the code as a tuple. The range of TOA densities for the program to randomly choose from so as to vary spacing between TOAs between simulated systems. 

\texttt{span} - Optional input, defaults to '200, 700' (days). The range of total time spans to choose from in units of days, where the comma separated pair is a string that will be interpreted in the code as a tuple

\begin{figure}[ht]
\plotone{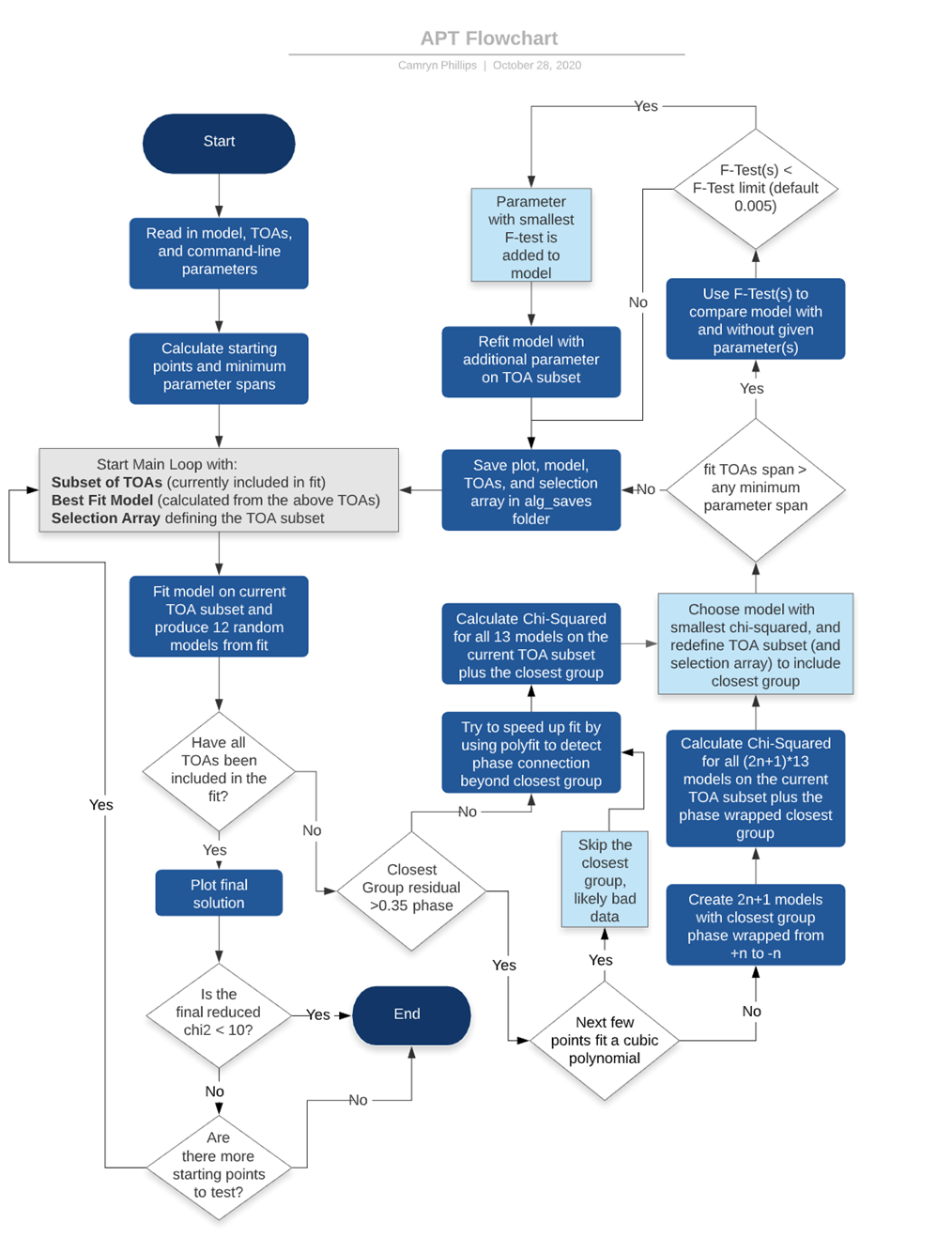}
\caption{Flowchart of APT's logic, showing the iterative process of fitting, adding TOAs and/or fit parameters, and checking for phase wraps.   \label{flowchart}}
\end{figure}

\end{document}